\documentclass[letterpaper,fleqn,usenatbib]{mnras}

\usepackage{enumerate}

\def\eg{{\it e.g.}}
\def\etal{{\it et al.}}

\def\Msun{M$_\odot$}

\def\pmb#1{\setbox0=\hbox{$#1$}%
  \kern-0.25em\copy0\kern-\wd0
  \kern.05em\copy0\kern-\wd0
  \kern-0.025em\raise.0433em\box0}
\def\spmb#1{\setbox1=\hbox{${\scriptstyle #1}$}%
  \kern-0.25em\copy1\kern-\wd1
  \kern.05em\copy1\kern-\wd1
  \kern-0.025em\raise.0433em\box1}

\long\def\Ignore#1{\relax}
\usepackage{color}
\definecolor{red}{rgb}{0.7,0.1,0.1}
\definecolor{blue}{rgb}{0.2,0.2,0.8}
\definecolor{green}{rgb}{0.1,0.6,0.1}

\usepackage{graphicx}	

\title[AGC 114905]{The ultra-diffuse galaxy AGC 114905 needs dark matter}

\author[Sellwood \& Sanders]
{J. A. Sellwood$^{1}$\thanks{E-mail:sellwood@as.arizona.edu}
and
R. H. Sanders$^{2}$\thanks{E-mail:sanders@astro.rug.nl}
\\
$^1$Steward Observatory, University of Arizona, 933 N Cherry Ave, Tucson AZ 85722, USA\\
$^2$Kapteyn Astronomical Institute, P.O. Box 800, NL-9700 AV Groningen, the Netherlands
}

\pubyear{2021}

\begin{document}
\label{firstpage}
\pagerange{\pageref{firstpage}--\pageref{lastpage}}
\maketitle

\begin{abstract}
Recent 21 cm line observations of the ultra-diffuse galaxy AGC~114905
indicate a rotating disc largely supported against gravity by orbital
motion, as usual.  Remarkably, this study has revealed that the form
and amplitude of the HI rotation curve is completely accounted for by
the observed distribution of baryonic matter, stars and neutral gas,
implying that no dark halo is required.  It is surprising to find a
DM-free galaxy for a number of reasons, one being that a bare
Newtonian disk having low velocity dispersion would be expected to be
unstable to both axi- and non-axisymmetric perturbations that would
change the structure of the disc on a dynamical timescale, as has been
known for decades.  We present $N$-body simulations of the DM-free
model, and one having a low-density DM halo, that confirm this
expectation: the disc is chronically unstable to just such
instabilities.  Since it is unlikely that a galaxy that is observed to
have a near-regular velocity pattern would be unstable, our finding
calls into question the suggestion that the galaxy may lack, or have
little, dark matter.  We also show that if the inclination of this
near face-on system has been substantially overestimated, the
consequent increased amplitude of the rotation curve would accommodate
a halo massive enough for the galaxy to be stable.
\end{abstract}

\begin{keywords}
galaxies: spiral ---
galaxies: evolution ---
galaxies: structure ---
galaxies: kinematics and dynamics ---
\end{keywords}


\section{Introduction}
\label{sec.intro}
Ultra diffuse galaxies (UDGs), which may comprise a distinct class of
large galaxies having very low surface brightness \citep{Imp88,
  Con18}, have recently attracted considerable attention.  These
objects are found in various environments ranging from large galaxy
clusters to voids and may be gas-rich or gas free -- facts that may
indicate various formation and evolution mechanisms.  Kinematic
studies of at least one gas-free system have implied very low
dynamical mass, consistent with the observed stellar mass, and it has
been suggested that these systems may be entirely free of dark matter
\citep{vDok19}.

More recently, 21 cm line studes of neutral hydrogen in several gas
rich UDGs have revealed similar atypical properties \citep{MP19}: it
appears that several do not lie on the well-established baryonic
Tully-Fisher relation \citep[][BTFR]{McG12}, in the sense that they
are too slowly rotating for their observed baryonic mass. One
particular galaxy, AGC~114905, is of comparable size to the Milky Way
but has a rotational velocity a factor 10 times lower. This object has
been studied in detail via 21 cm line observations by
\citet[][hereafter MP22]{MP22}.  They fitted a differentially rotating
inclined disc to the observed data cube, deriving a rotation curve
that is apparently consistent with the attraction of the observed
distribution of stars and gas only.  Basically, the detected baryonic
matter, mostly gas which has no uncertainty in the stellar
mass-to-light ratio, gives rise to a rotation curve that very closely
matches that fitted to their data.

Disc galaxies that do not appear to be embedded in a dark matter halo
would be remarkable for two principal reasons.  First, the modern
$\Lambda$CDM theory of galaxy formation \citep[reviewed by][]{SD15}
supposes that galaxies form in pre-existing dark matter concentrations
\citep[but see also][]{More22}.  Second, rotationally-supported
galaxies lacking DM halos and a central bulge have long been known
\citep{MPQ, Hohl71} to be globally unstable to the formation of a bar,
which provided one of the original motivations for DM halos
\citep{OP73}.  The history of these issues is discussed in detail by
\citet{sand}.  Note that a DM free galaxy would also be clearly
inconsistent with the MOND paradigm \citep{Milg83} which requires
every relatively isolated galaxy to have an apparent ``halo''.  These
considerations account for the intense interest in UDGs.

One concern with the study by MP22 is that their estimated inclination
of the disc plane is $i \sim 32^\circ$, which they based on the
apparent ellipticity of the outer neutral hydrogen disc.  An accurate
estimate of $i$ is difficult to determine when the galaxy is close to
face on, causing most practitioners of rotation curve fitting to
exclude such cases; for example \citet{McG12} limits his study to
galaxies having $i > 45^\circ$.  This issue was addressed by
\citet{Read16b}, who applied the usual observational techniques to
simulated galaxies, and concluded that dwarf galaxies with $i \la
40^\circ$ may have large inclination errors.  The dwarf galaxy IC~1613
is a case in point; \citet{Lake89} estimated $i \simeq 39^\circ$ and
concluded that the galaxy was both deficient in dark matter and
inconsistent with MOND.  \citet{Read16b} showed IC~1613 to be an
outlier from the BTFR, and suggested the inclination had been
overestimated; indeed \citet{Oman16} argued, from an indirect
interpretation of the stellar velocity dispersion, for $i \simeq
20^\circ$.  Furthermore, the paper \citet{Bani22} appeared while we
were waiting for the referee to report on this paper, and those
authors argue that the inclination of AGC~114905 may be as low as
$\sim 12^\circ$ in order that it be consistent with the BTFR and with
MOND.

Since a fully self-gravitating disc galaxy is expected to be gobally
unstable, we have constructed $N$-body realizations of the models of
AGC~114905 fitted by MP22 to their data and evolved them to examine
their stability.  Full details of the models and our methods are
given in the following section. 

Implicit in the study presented here are the assumptions that the
galaxy is undisturbed, in a long-lived state, and that the observed
rotation pattern reflects centrifugal balance.  Although the disc has
not had many dynamical times to settle (the orbit period at the disc
edge, $R=10\;$kpc, is $\sim 2.6\;$Gyr), the apparent regularity of the
observed velocity field gives some credence to these assumptions but,
were they not to hold, there is little that can be deduced from
stability considerations.

\begin{figure}
\includegraphics[width=.99\hsize,angle=0]{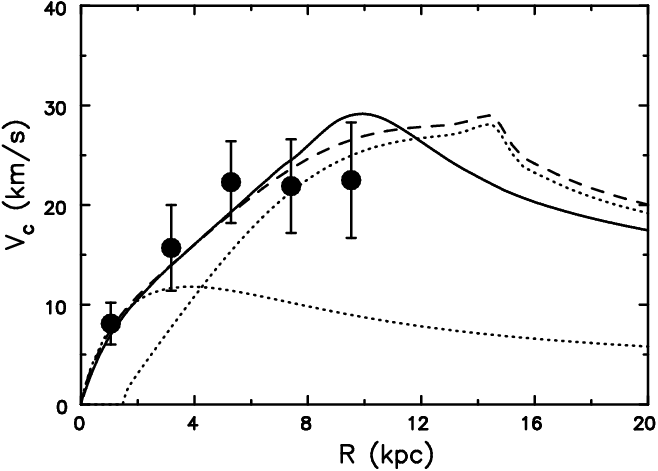}
\caption{The data points with error bars show the circular speed in
  AGC~114905 measured by MP22 for an adopted inclination of
  32$^\circ$.  The dashed curve to $R\la 10\;$kpc shows the circular
  speed in the no-DM model proposed by MP22 that is the sum, in
  quadrature, of the dotted curves, which are the contributions of the
  two disc components.  The solid curve is the circular speed measured
  directly from the attaction of the particles in our simulation, see
  the text for more details.  Note that the separate calculated
  rotation curves assume the surface density of the gas was tapered to
  zero near $R=15\;$kpc for the dashed curve and more gently and
  farther in for the solid curve.  The solid and dashed curves are in
  good agreement inside $R\sim 8\;$kpc, but the solid curve exceeds
  that expected from the extended gas disc for $8 \leq R \leq
  10\;$kpc.}
\label{fig.IRC}
\end{figure}

\section{Technique}
\label{sec.methods}

\subsection{No DM model}
\label{sec.model}
Our first simulation is of a model without DM, whose properties are
given in \S4.1 of \citet{MP22}.  These authors present neutral
hydrogen observations of the galaxy from which they derive a
deprojected circular speed that rises gently from zero at the centre
to $\sim 23\;$km~s$^{-1}$ at $R=10\;$kpc.

MP22 fit the projected star light distribution as a round exponential
disc and, adopting $M/L=0.47$ in the $r$-band, give the surface mass
density as
\begin{equation}
\Sigma_{\rm stars}(R) = {M_d \over 2\pi R_d^2} \exp(-R/R_d).
\label{eq.surfds}
\end{equation}
Here $R_d = 1.79\;$kpc is the disc scale length and $M_d = (1.3\pm0.3)
\times 10^8\;$\Msun\ is the mass of the notional infinite disc, the
uncertainty reflecting that in the $M/L$.  We taper the stellar disc
surface density to zero over the range $3.5R_d \leq R \leq
4R_d$. There are no data (at the time of this writing) to indicate the
stellar velocity dispersion in this galaxy, so we adopt $Q_{\rm stars}
\ga 1.5$ \citep{To64}.  The Jeans equations in the epicyclic
approximation \citep{BT08} are inadequate to construct a reasonable
equilibrium for the star particles because the radial velocity
dispersion is $\ga 7\;$km~s$^{-1}$ near the disc centre which is
comparable to the observed circular speed.  We therefore derive a 2D
distribution function (DF) for the disc using the procedure described
by \citet{Shu69}.  As the vertical density profile of the disc is also
unknown, we adopt a Gaussian profile with a scale $0.1R_d$, and use
the 1D Jeans equation to set up the appropriate vertical velocity
dispersion, which results in a slightly flattened velocity ellipsoid
at all radii.

\begin{figure}
\includegraphics[width=.99\hsize,angle=0]{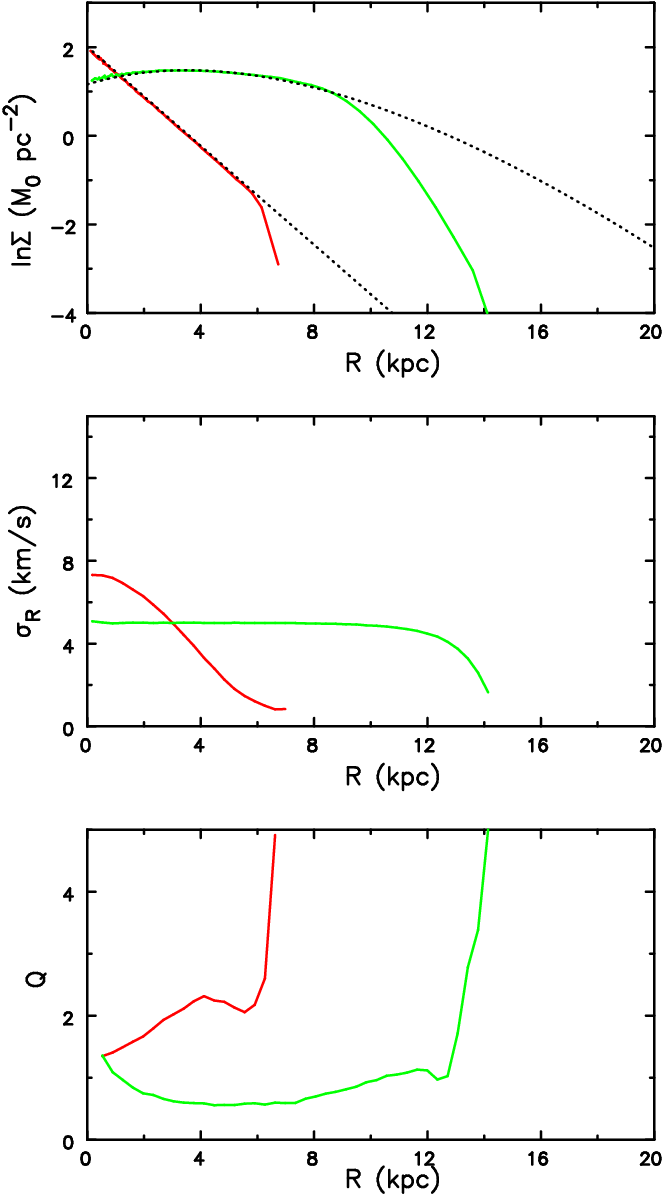}
\caption{The solid lines in each panel are measured from the start of
  the no-DM matter simulation.  The top panel shows the tapered
  surface densities of star particles (red) and of gas particles
  (green), while the dotted lines are the untapered functions
  (eqs.~\ref{eq.surfds} and \ref{eq.surfdg}).  The middle panel shows
  the radial velocity dispersion of the particles, while the bottom
  gives the stability parameter, $Q$, for each component.}
\label{fig.iprops}
\end{figure}

\begin{table}
\caption{Default numerical parameters} 
\label{tab.DBHpars}
\begin{tabular}{@{}ll}
Grid points in $(r, \phi, z)$ & 175 $\times$ 256 $\times$ 125 \\
Grid scaling & $R_d= 5$ grid units \\
Vertical spacing & $\delta z = 0.04R_d$ \\
Active sectoral harmonics & $0 \leq m \leq 8$ \\
Softening length & $R_d/20$ \\
Number of star particles & $10^6$ \\
Number of ``gas'' particles & $10^7$ \\
Time-step & $(R_d^3/GM)^{1/2}/80 = 1.24\;$Myr \\
\end{tabular}
\end{table}

\begin{figure*}
\includegraphics[width=.4\hsize,angle=270]{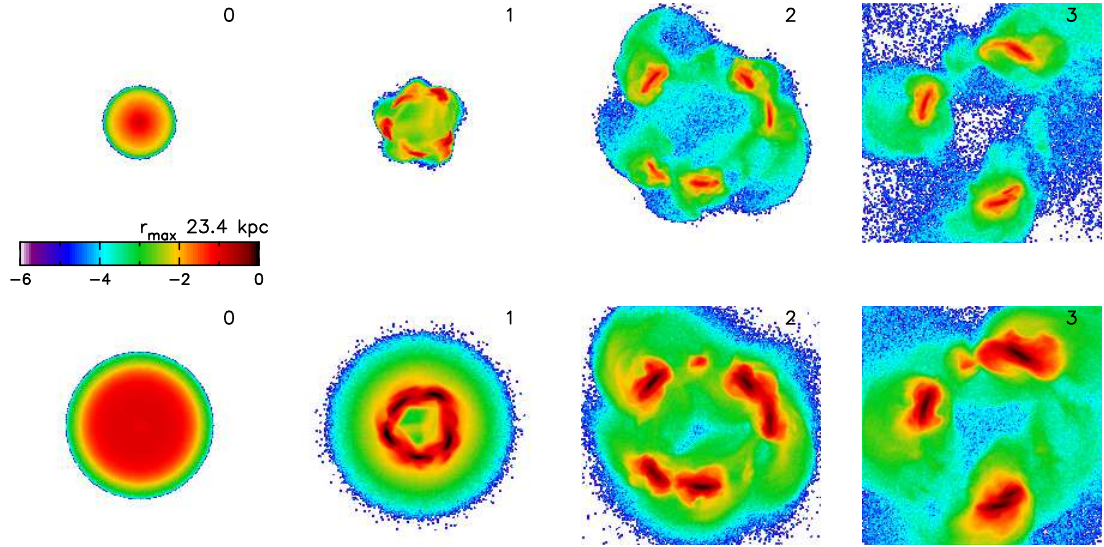}
\caption{The evolution of the star component, top row, and gas
  component, bottom row, in the model with no dark matter.  The colour
  scale indicates the logarithm of the projected particle density,
  times are in Gyr and the box is 46.8~kpc on a side.}
\label{fig.proj1}
\end{figure*}

MP22 fit the observed gas surface density, corrected for helium
content, with a function of the form
\begin{equation}
\Sigma_{\rm gas}(R) = \Sigma_{0, \rm gas} \exp(-R/R_1)(1 + R/R_2)^\alpha,
\label{eq.surfdg}
\end{equation}
where the central surface density of the gas $\Sigma_{0, \rm gas} =
3.2\;$\Msun~pc$^{-2}$, $R_1 = 1.11\;$kpc, $R_2 = 16.5\;$kpc, and
$\alpha = 18.04$.  The gas disc extends to $R \sim 10\;$kpc, and has a
total mass of $(1.29\pm0.19)\times10^9\;$\Msun; it is therefore both
more extensive and ten times more massive than the stellar disc.  MP22
point out that the attraction of the two disc components gives rise to
a rotation curve, the dashed line in Figure~\ref{fig.IRC}, that is in
reasonable agreement with the observed velocities.  MP22 measure the
velocity dispersion of the gas component to be close to $\sim
5\;$km~s$^{-1}$ at most radii and $\sim 8\;$km~s$^{-1}$ at their
innermost point.  We adopt an approximately isotropic velocity
dispersion of $5\;$km~s$^{-1}$ at all radii.

In most of our simulations, we model the gas disc as a collection of
collisionless $N$-body particles, but we have tried the SPH option in
{\sc Gadget-2} in one case (see \S\ref{sec.SPH}).

Once again, we select ``gas'' particles from a 2D DF created by Shu's
procedure in order to build an equilibrium disc model, although in
this case the outer edge of the disc presented an extra difficulty.
The function (eq.~\ref{eq.surfdg}) fitted to the disc surface density
peaks near $R \sim3\;$kpc and declines gently to beyond $R = 10\;$kpc
with no outer truncation.  As is well known \citep{To63, Ca83},
introducing a sharp truncation to a disc creates a blip in the central
attraction, causing the circular speed to rise steeply to the
truncation radius and to decline more quickly than the Keplerian rate
just beyond.  MP22 report that the observed circular speed stays flat
over the last few measured points, as shown in Figure~\ref{fig.IRC},
revealing no kinematic signature that the gas disc has an edge, even
though their Fig.~1 hints that the HI disc ends quite sharply near
$R\sim 10\;$kpc.  We were told (Fraternali, private communication)
that this apparent edge is probably an artifact introduced by a mask
in their data reduction procedure, and the gas disc may extend
smoothly to larger radii.

Since there are no kinematic data for $R\ga 10\;$kpc, we prefer to
truncate the gas disc, even though it creates a feature in the central
attraction that is not observed.  However, it is numerically
inconvenient to find a DF and to select particles in a potential that
causes a blip in the rotation curve near the disc edge.  We therefore
tapered the surface density of the gas disc over the radial range
$10.7 < R < 14.3\;$kpc in order to moderate the effect of truncation,
although a broader and more gentle bump in the circular speed curve
remains, as shown by the solid curve in Figure~\ref{fig.IRC}.  In
order to avoid numerical problems, we computed the DF and selected
particles in a potential derived from a gas disc having the taper
shifted to a larger radius, which made it easier to assign orbital
velocities to the gas particles, although the actual surface density
of the selected particles was tapered as noted above.  The attraction
in the outer disc ($8\leq R \leq 10\;$kpc) is stronger in the
simulation (solid curve) than assumed when selecting particles from
this DF (dashed curve), which causes the gas particles to have a
somewhat lower orbital speed than needed for centrifugal balance.  The
disc inside 8~kpc, which contains most of the mass, is in proper
equilibrium, and we do not expect the mild imbalance in the outer
part, which causes a slight initial contraction, to affect the overall
stability.

The tapered surface density profiles are shown in the top panel of
Figure~\ref{fig.iprops}, and the other two panels present properties
of this model measured from the particles at the start of the
simulation.  Note that $Q_{\rm gas}$ in the bottom panel, which is
determined by the measured velocity dispersion, surface density, and
rotation curve, is less than unity over a broad radial range.

\subsection{Numerical methods}
The particles in our simulations move in a 3D volume that is spanned
by a cylindrical polar mesh.  The self-gravitational attractions are
calculated at grid points and interpolated to the position of each
particle.  A full description of our numerical procedures is given in
the on-line manual \citep{Sell14} for the {\sc Galaxy} code and the
source is available for download.  Table \ref{tab.DBHpars} gives the
values of the numerical parameters adopted for the simulations
presented in this paper.

We have verified that the results we obtain are insensitive to the
numbers of particles employed, whether the particles are confined to a
plane or move in 3D, and also when evolved using the $N$-body option of
{\sc Gadget-2} \citep{Spr05}.  We describe below (\S\ref{sec.SPH}) the
behaviour of a model in which we turned on SPH for the gas particles.

\begin{figure}
\includegraphics[width=.9\hsize]{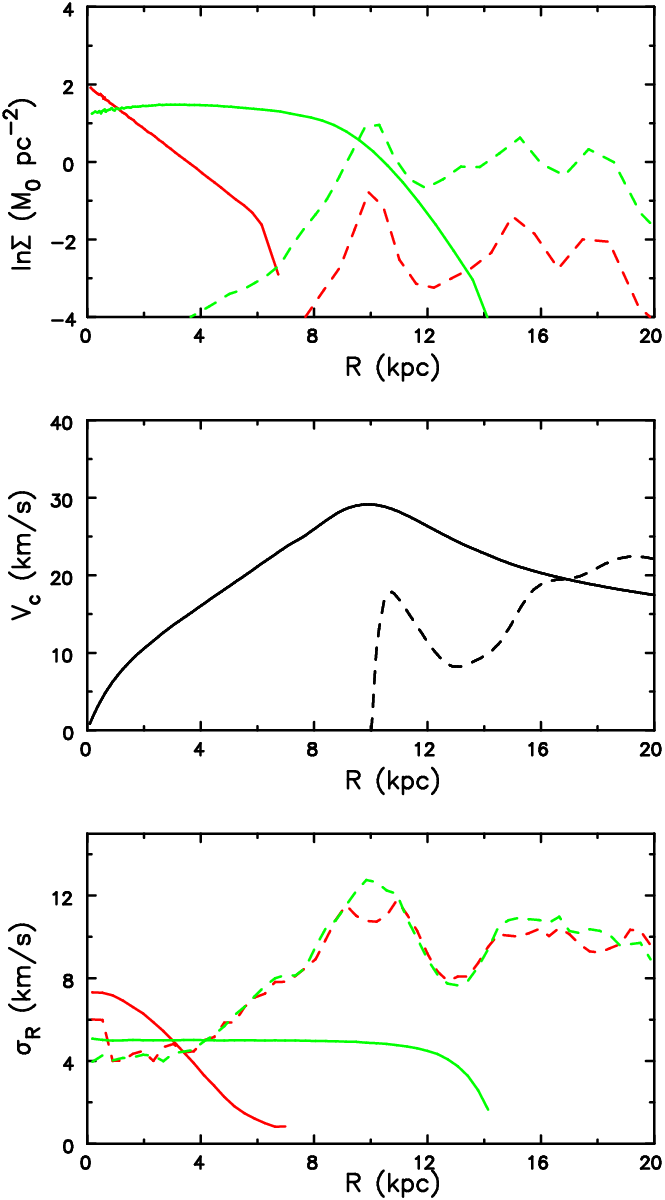}
\caption{The initial (solid curves) and final (dashed curves) of the
  azimuthally averaged surface density (top panel) and radial veocity
  dispersion (bottom panel) in the model shown in
  Figure~\ref{fig.proj1}.  The colours are red for stars and green for
  gas, as in Figure~\ref{fig.iprops}.  The middle panel gives the
  initial and final rotation curves.}
\label{fig.m1-final}
\end{figure}

\section{Results}
\label{sec.res1}
We evolved the model presented in \S\ref{sec.model} for 3~Gyr, and a
few snapshots are shown in Fig.~\ref{fig.proj1}.  The evolution of the
stellar component is illustrated in the top row and the gas particles
in the bottom row.  The more massive gas component, which initially
had $Q_{\rm gas}<1$ over a broad radial range (bottom panel of
Fig.~\ref{fig.iprops}), and thus was axisymmetrically unstable
\citep{To64}, formed a ring at first which then fragmented.  Note the
stellar disc had $Q_{\rm stars}>1$ everywhere but the attraction of
the gas particles caused it to behave in a similar manner.

Although the initial model was in equilibrium, the rapid growth of
instabilities profoundly changed the structure of both discs, as shown
in Figure~\ref{fig.m1-final}.  Both discs expanded leaving a very low
central density (top panel) and, because angular momentum is
conserved, the orbital speeds of the particles decreased, while random
motion was greatly increased (bottom panel).  We stopped the
simulation when the rate of change of these properties was beginning
to slow.  The middle panel shows a ``rotation curve'' at the last
moment that differs profoundly from that observed.

\subsection{Gas physics}
\label{sec.SPH}
We resimulated this model using {\sc Gadget-2} initially as $N$-body
particles, obtaining very similar results as with {\sc Galaxy}.  We
then attempted to apply the SPH option of {\sc Gadget-2} for the gas
component.  In this case, we adopted an isothermal equation of state
for the gas with a sound speed of $5\;$km~s$^{-1}$ and no star
formation or feedback.  We found that the gas disc began to fragment
even more rapidly than when employed collisionless particles, and the
gas clumps reached such a density that {\sc Gadget-2} stopped after
just $0.24\;$Gyr.  This well-known behaviour of dissipative gas with
self-gravity may possibly be inhibited by allowing star formation and
feedback, but we considered that implementing this option would be a
major digression that would be most unlikely to alter the conclusion
that the model is globally unstable.

\begin{figure}
\includegraphics[width=.9\hsize,angle=0]{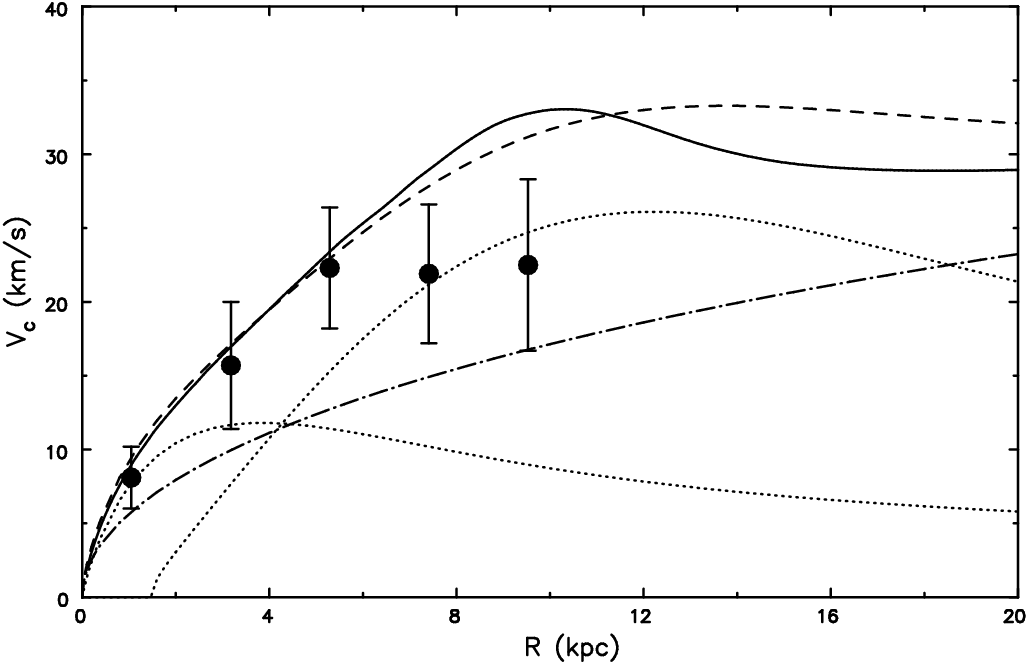}
\caption{The initial rotation curve of the ``Case 2'' model having the
  low-concentration DM halo.  The data points have been reproduced and
  the lines are the same as in Fig.~\ref{fig.IRC}, except that the
  dot-dashed line gives the contribution of the halo to the new total
  circular speed, marked by the dashed and solid curves.}
\label{fig.DMH}
\end{figure}

\begin{figure*}
\includegraphics[width=.4\hsize,angle=270]{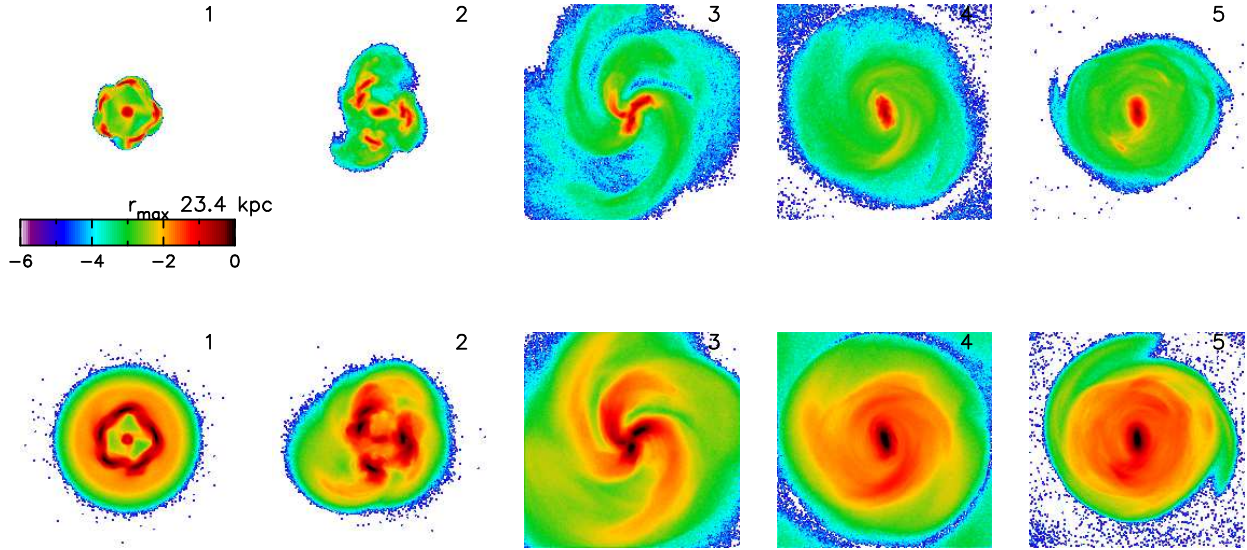}
\caption{As for Fig.~\ref{fig.proj1}, but for the model having a
  low-concentration DM halo.  Note that the times of the snapshots
  also differ.}
\label{fig.proj2}
\end{figure*}

\subsection{Including a dark matter halo}
\label{sec.DM}
Although the observed velocities in AGC~114905 could be accounted for
without including a DM component, MP22 noted that some dark matter
content should be expected.  They therefore presented two models that
included dark matter halos having {\sc coreNFW} \citep{Read16a} density
profiles.  They restricted the allowed parameters of the halo to
values in the cosmologically expected range in their Case 1, but were
unable to fit circular speeds as low as observed.  So they removed the
restriction on the concentration parameter and present a model, their
Case 2, that has $c=0.3$.  Adding a halo of mass $M_{200} =
10^{10}\;$\Msun\ and this low concentration, while retaining the
previously fitted disc densities for the stars and gas, resulted in a
rotation curve with higher amplitude, but perhaps consistent with
their observed values within the errors, as shown
Figure~\ref{fig.DMH}.  MP22 acknowledge that the low concentration of
this halo is well below the range expected for $\Lambda$CDM halos.

We have therefore created an $N$-body realization of the two-component
disc in the rigid gravitational potential of their Case 2 DM halo
component.  The different gravitational field required us to rederive
new DFs for both discs.  We kept the properties of both discs as
before, except that we thickened the gas layer in order that the
vertical velocity dispersion was very similar to the 5~km~s$^{-1}$
assigned to the radial component to make the velocity ellipsoid more
nearly isotropic.  (Epicyclic motions require the azimuthal dispersion
to differ from the radial component.)  Since the circular speeds are
higher, the epicyclic frequency is also increased by the DM halo,
slightly raising the initial $Q_{\rm gas}$, although it is still less
than unity over the range $2<R<8\;$kpc.

\begin{figure}
\includegraphics[width=.9\hsize,angle=0]{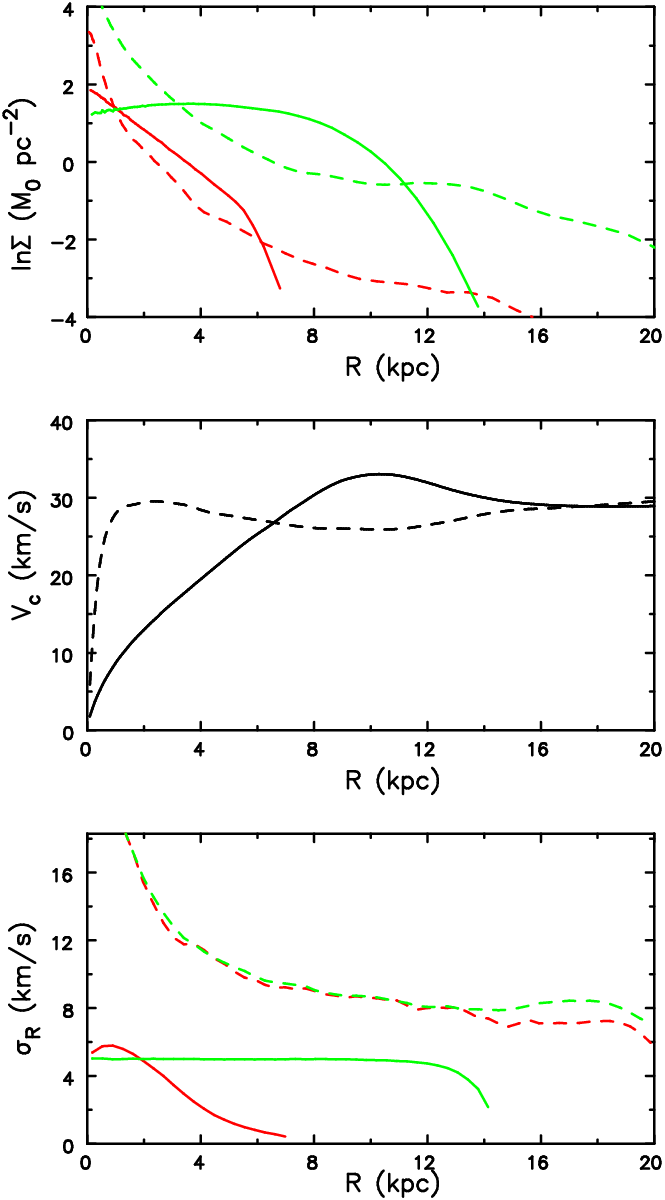}
\caption{Same as Figure~\ref{fig.m1-final} but for the model shown in
  Figure~\ref{fig.proj2}.}
\label{fig.m2-final}
\end{figure}

\begin{figure*}
\includegraphics[width=.38\hsize,angle=270]{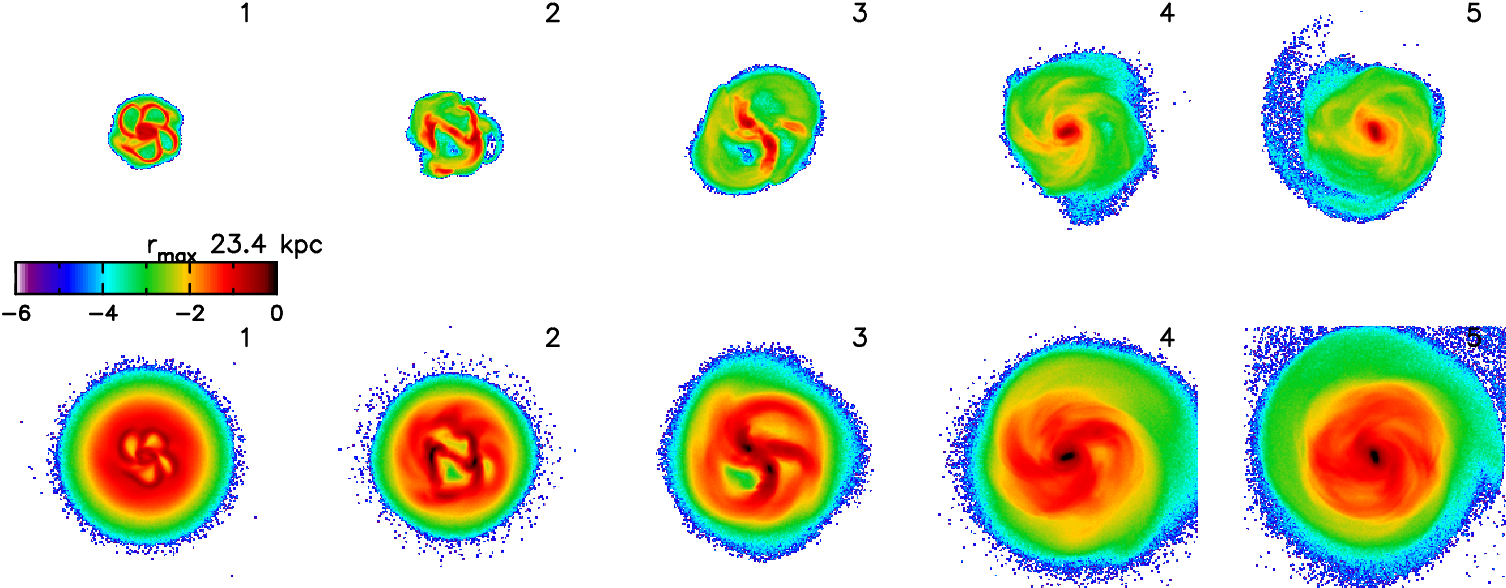}
\caption{As for Fig.~\ref{fig.proj1}, but for a model in which we
  reduced the estimated inclination to $20^\circ$.}
\label{fig.projinc1}
\bigskip
\includegraphics[width=.38\hsize,angle=270]{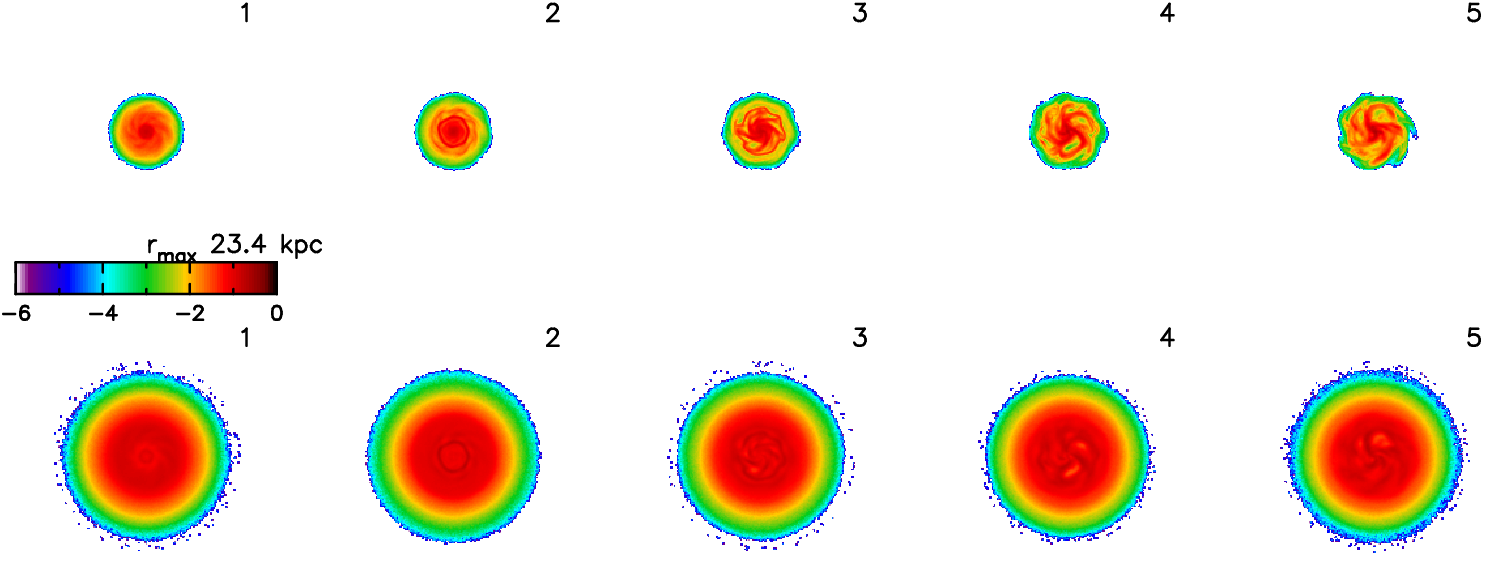}
\caption{As for Fig.~\ref{fig.projinc1}, but for the estimated
  inclination of $15^\circ$.}
\label{fig.projinc2}
\end{figure*}

The evolution of this model is presented in
Figure~\ref{fig.proj2}. Since this simulation evolved more slowly and
less violently, we ran it for longer.  The evolution was also changed
in other respects: an almost axisymmetric instability was again the
first to develop, but non-axisymmetric features were already prominent
by $t=1\;$Gyr.  The order of rotational symmetry of these features
decreased over time and both the stellar and gas particles developed a
bar by $t\sim3\;$Gyr that persisted to the end.

Figure~\ref{fig.m2-final} compares the initial and final properties of
this model.  The discs did not spread by quite as much (top panel) and
their central densities rose rather than declined, unlike in first
simulation, causing a profound change to the rotation curve (middle
panel). The level of random motion was also greatly increased by the
end of the simulation (bottom panel).  Once again, we consider these
changes to be extensive enough to indicate that the initial model was
unacceptable as a long-lived state.

We treated the DM halo in this simulation as a rigid component.
Previous work \citep{Ath02, Sell16} has shown that realizing the halo
with mobile particles causes the disc to be more unstable.

\begin{figure*}
\hbox to \hsize{
\includegraphics[width=.45\hsize]{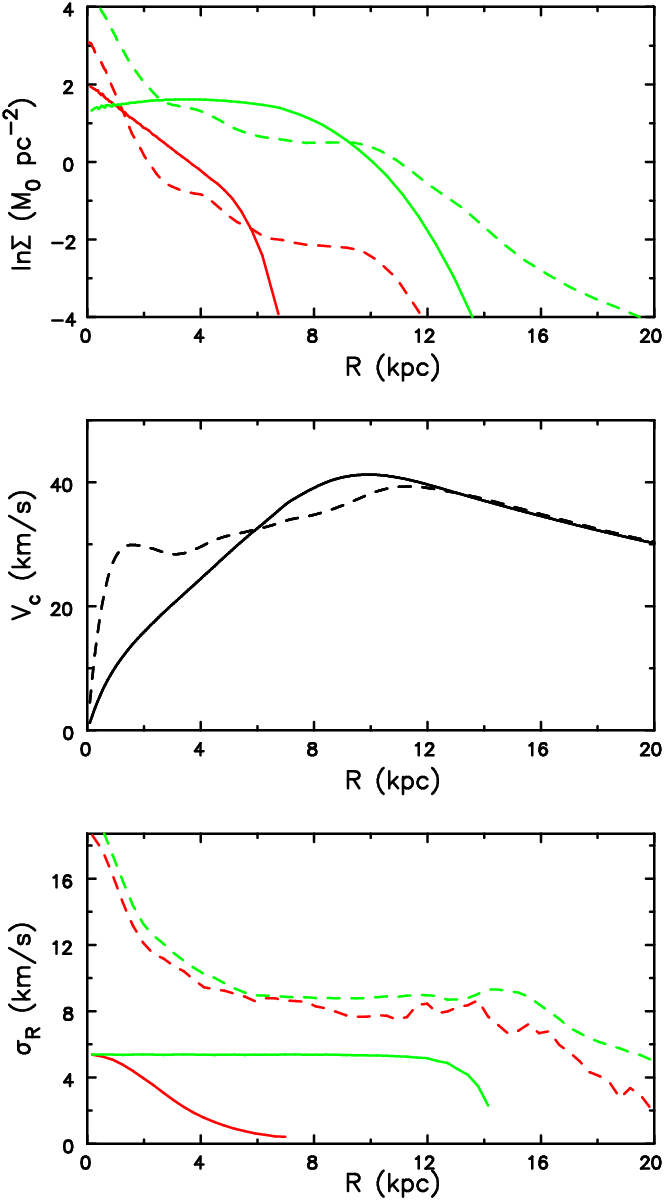} \hss
\includegraphics[width=.45\hsize]{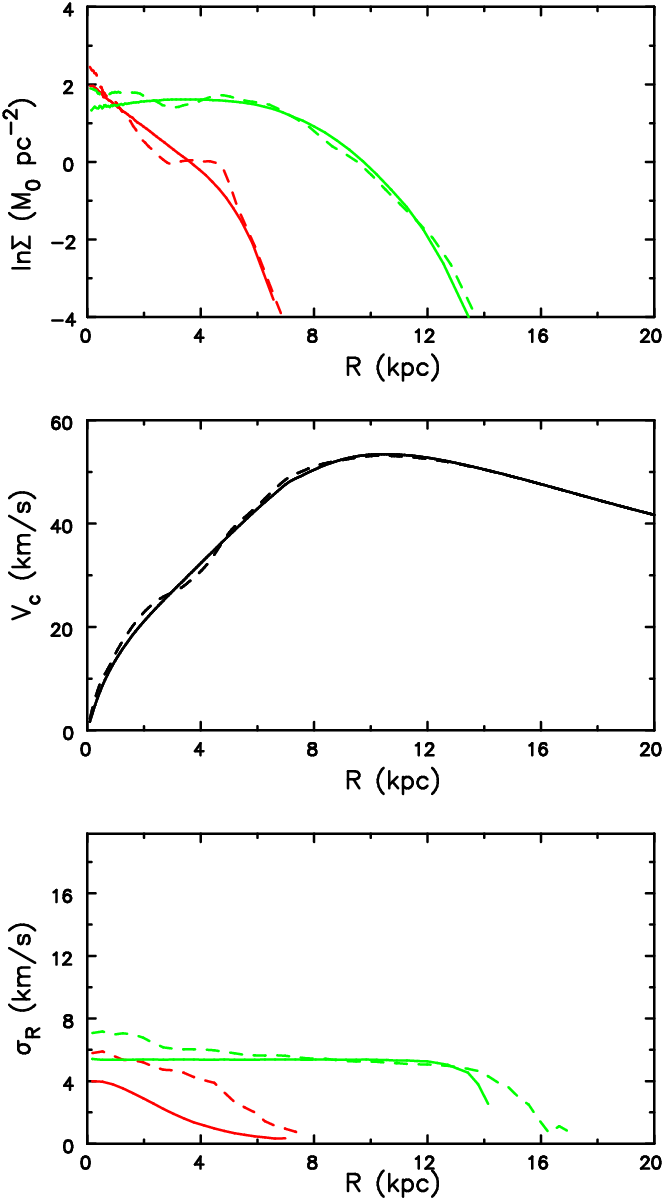}}
\caption{Same as Figure~\ref{fig.m1-final} but for the models shown in
  Figures~\ref{fig.projinc1} on the left and \ref{fig.projinc2} on the
  right.}
\label{fig.m20-final}
\end{figure*}

\section{Searching for a stable model}
The gas and stars in the galaxy AGC~114905 manifest some density
variations, but it is very unlikely that MP22 observed it at a moment
just before the discs disrupted due to instabilities.  Therefore we
have attempted to find a different stable, or slowly evolving model
that matches their data.  However, the stubborn fact is that the gas
disc is alone massive enough to account for most of the inferred
circular speed, and has $Q<1$ over a broad radial range.  These
properties inevitably imply instability.

Note $Q \equiv \sigma_R \kappa / (3.36 G\Sigma)$ \citep{To64}, with
$\kappa$ being the usual epicyclic frequency \citep{BT08}.\footnote{We
  prefer this definition for a turbulent, clumpy gas rather than $Q
  \equiv c \kappa / (\pi G\Sigma)$ that applies to a smooth barotropic
  fluid having a sound speed $c$, but the difference is minor.}  Thus
$Q$ could be higher if either $\sigma_R$ has been underestimated, or
the circular speed has been underestimated, implying a higher value
for $\kappa$, or the surface density of the gas has been
overestimated, or any combination of these factors.

MP22 report an HI mass, without correction for helium, of $M_{\rm HI}
= (9.7\pm1.4) \times 10^8\;$\Msun, which is lower than $M_{\rm HI}
\approx 1.3 \times 10^9\;$\Msun\ implied by the ALFALFA flux; the
difference can probably be ascribed to missing short baselines in the
interferometric data.  Thus these data would seem to preclude the idea
that the tolerance in the HI mass may be consistent with a
significantly lower surface density for the gas.

Although the stellar disc mass $M_* = (1.3\pm0.3) \times
10^8\;$\Msun\ is less precisely known, its mass is merely $\sim 10\%$
of the mass of the gas disc and correspondingly less important to
stability.  We found that the evolution of a DM-free simulation was
little affected by a 1-$\sigma$ reduction in the mass of the stellar
disc.

These considerations suggested two distinct strategies find stable,
long-lived models.

\subsection{Changing the disc inclination}
\label{sec.incl}
One factor that could change the stability is a reduction to the
adopted inclination of the disc, which may have been overestimated,
as discussed in the introduction.  If it were in fact lower than the
$\sim32^\circ$ preferred by MP22 the rotation curve amplitude would
be higher, which would in turn allow a more massive halo, as MP22 note
in their \S5.4.  We have therefore tried a series of simulations to
find the maximum inclination at which the stellar and gas discs,
having the properties described in \S2 above, would be stable, or
slowly evolving.

We do not need to construct a halo model to make these tests, since we
can add to the radial forces from the particles the central attraction
of whatever rigid mass is needed to account for the difference between
the rotation curve that arises from the baryonic matter, which we
already know, and the scaled up rotation curve.  The radial density
profile of this implied halo has a uniform core inside $\sim100\;$pc
and the density declines as $r^{-\alpha}$ over the range $2<r<8\;$kpc,
with $\alpha \sim 0.9$.  This is quite different from the profile
expected in $\Lambda$CDM galaxy formation theory.

Figures~\ref{fig.projinc1} \& \ref{fig.projinc2} present the results
from two simulations in which we allowed higher circular speeds
through supposing the inclination to have been been overestimated.
Figure~\ref{fig.projinc1} shows the effect of increasing the circular
speed shown in Fig.~\ref{fig.IRC} by a factor of 1.5, which implies an
inclination of $i \sim20^\circ$ instead of $i \sim30^\circ$ preferred
by MP22.  Since the contributions from the stellar and gaseous discs
are unchanged, and the rotation curve is simply scaled up, the implied
DM mass to $R\sim 10\;$kpc is 1.23 times the combined masses of the
two discs.  As may be seen, this DM component weakens the
instabilities in the two discs, although they continued to develop; a
bar formed after 4~Gyr, which weakened by $t\sim5\;$Gyr.

The final properties of this model are shown in the left-hand panels
of Figure~\ref{fig.m20-final}.  The changes in density (top panel) and
velocity dispersion (bottom panel) resemble those in
Figure~\ref{fig.m2-final}, and the changes occurred over a similar
time period.  The middle panel reveals that the change to the rotation
curve is a little less significant than in Figure~\ref{fig.m2-final},
probably because of the substantial contribution from the rigid halo.

We doubled the amplitude of the rotation curve for the simulation
shown in Figure~\ref{fig.projinc2}, implying an inclination of the
galaxy $i \sim 15^\circ$ and a DM halo mass 2.9 times the combined
mass of the two discs inside $R\sim 10\;$kpc. In this case, neither
disc evolved significantly as shown in the right-hand panels of
Figure~\ref{fig.m20-final} and the rotation curve was little changed
by the end.

From these results, we learn that the apparently puzzling instability
of the Case 2 model for AGC~114905 proposed by MP22, goes away if the
inclination of the galaxy to the line of sight is reduced from their
adopted $i\approx30^\circ$ to $\sim 15^\circ$.  The higher circular
speed in this model allows room for a DM mass $\sim3$ times the
combined baryonic mass of the stellar and gaseous discs, thereby
slowing the evolution and removing the challenge presented by their
model.  If the true inclination were in the range $10^\circ \la i \la
12^\circ$, this object would place clearly on the baryonic
Tully-Fisher relation \citep{McG09}.

MP22 determine the inclination of the disc in AGC~114905 from the
apparent ellipticity of the gas disc, which they assume to be
intrinsically round.  Decreasing their fitted value of $i=32^\circ$
therefore implies that the gas disc may be mildly {\it non-axisymmetric}.
Indeed, the map of the HI surface density \cite[][their Fig.~1]{MP22}
shows some evidence of this; the axis ratio of the surface density
contours is not constant and appears to change position angle with
radius.  This is not consistent with an axisymmetric disc seen in
projection.  Deeper HI observations of this, and of other gas-rich
UDGs, may shed further light on this issue.

\subsection{Stretching the data}
\label{sec.stretch}
Rather than pinning all the blame on a error in the estimated
inclination, Pavel Mancera Pi\~na and Filippo Fraternali (private
communication, 2022), suggested that all the observational
uncertainties may have conspired together to favour instability, and
that a stable model could be consistent with the data, within the
measured uncertainties.  They therefore asked us to examine the
stability of a model with a {\sc coreNFW} halo that was more massive
than that in their Case 2 model that we examined above
(Figures~\ref{fig.proj2} and \ref{fig.m2-final}).

\begin{figure}
\includegraphics[width=.9\hsize,angle=0]{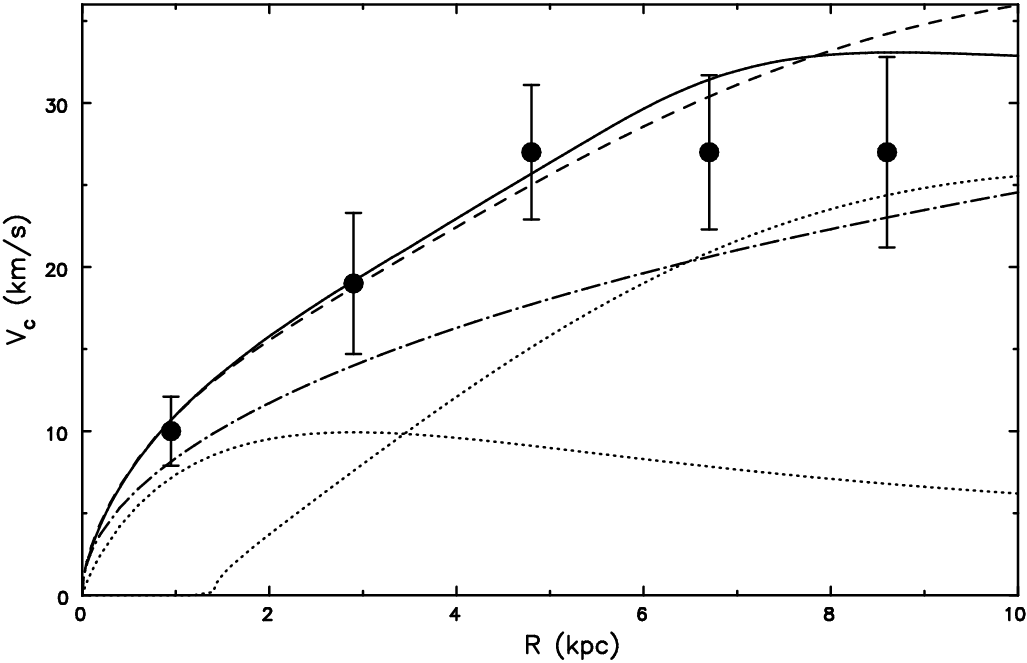}
\caption{The initial rotation curve of Case 3, the model described in
  \S\ref{sec.stretch} in which the parameters have been stretched.
  The lines are the same as for Case 2 in Fig.~\ref{fig.DMH}, and the
  observed circular speeds with error bars have been changed from
  those above to reflect the revised inclination and distance.}
\label{fig.DMH2}
\end{figure}

To construct their new model, which we refer to as Case 3, they first
reduced the estimated distance to AGC~114905 from 76~Mpc to 66~Mpc, a
2-$\sigma$ change, that reduced the masses, and sizes, of the discs.
They also reduced the M/L of the stars by 1-$\sigma$.  The new discs
have the same functional forms as given in equations~\ref{eq.surfds}
and \ref{eq.surfdg}, but with revised values of the parameters:
$\Sigma_{0, \rm stars} = 6.3\;$\Msun/pc$^{-2}$ and $R_d = 1.35\;$kpc
for the stellar disc and $\Sigma_{0,\rm gas} = 3.4\;$\Msun/pc$^{-2}$
and $R_1 = 1.00\;$kpc, $R_2 = 14.82\;$kpc, with unchanged
$\alpha=18.04$ for the gas disc.  The mass of the stellar disc is
$10^8\;$\Msun\ and the gas mass inside 10~kpc is
$1.04\times10^9\;$\Msun.  They also reduced the inclination by
2-$\sigma$ to $26^\circ$, and included the most massive halo that
would not cause the circular speed to exceed the observed values by
more than twice the measured uncertainty.  The parameters of the {\sc
  coreNFW} halo they adopt are: $M_{200} = 10^{10.45}\;$\Msun,
$c_{200} = 0.8$, $n=0.05$, and $r_c=4.59\;$kpc.

They finally suggested that the velocity dispersion in the gas could
be raised within the tolerances of their measurements, although they
concede that the low signal-to-noise of their data makes the
dispersion, and especially its uncertainty, difficult to estimate.  We
have therefore treated it as a free paramater and run three separate
models with $\sigma_{\rm gas} = 5$, 8, and 11~km~s$^{-1}$ at all
radii.  Note that we also made corresponding changes to the thickness
of the gas layer in order to keep a roughly isotropic distribution of
the velocities for the gas in each case.

We adopted a standard NFW halo with $M_{200} = 10^{10.45}\;$\Msun\ and
$c_{200} = 0.8$, and also kept the tapers of surface density the same
in disc scale lengths as before, so that our realizations of their
Case 3 model did not have exactly the rotation curve they proposed.
Our model is shown in Figure~\ref{fig.DMH2}, and compared with the
measured circular speeds, which have been adjusted for the revised
inclination and distance.  The DM mass enclosed within $10\;$kpc is
$1.40\times10^9\;$\Msun, which exceeds the total baryonic mass,
$1.14\times10^9\;$\Msun, interior to that radius.

\begin{figure}
\includegraphics[width=.9\hsize]{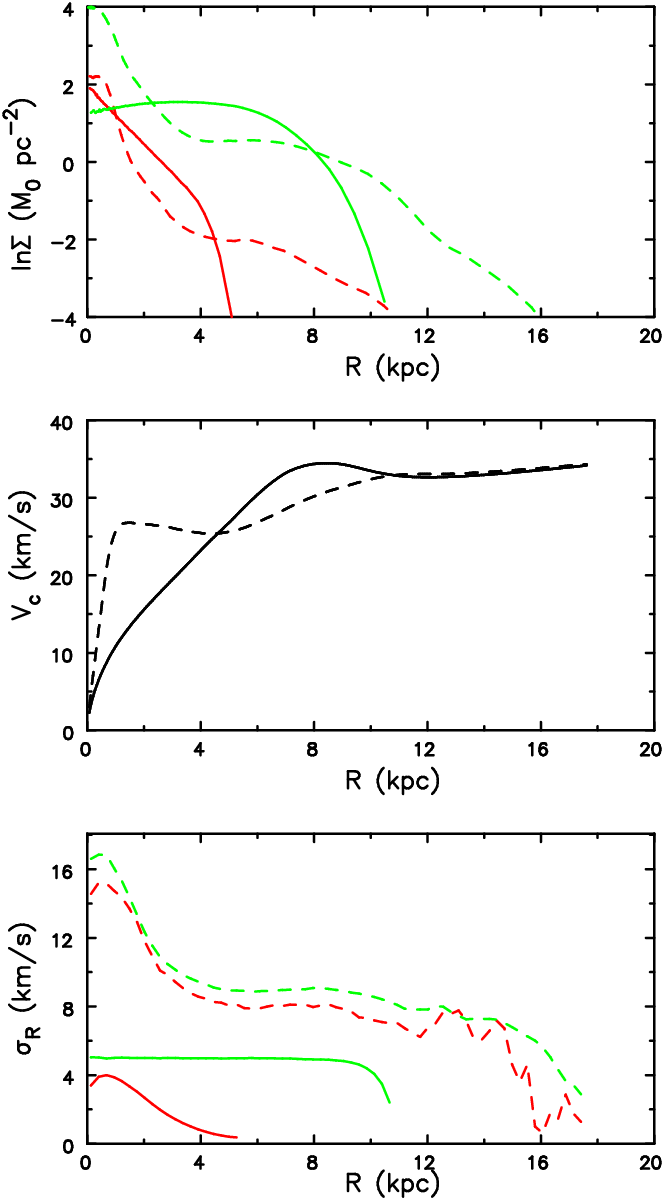}
\caption{Same as Figure~\ref{fig.m1-final} but for Case 3 with
  $\sigma_{\rm gas} = 5\;$km~s$^{-1}$, showing that the model is
  unstable.  Simulations having the same mass distribution but
  $\sigma_{\rm gas} \geq 8\;$km~s$^{-1}$ were stable.}
\label{fig.m3-5final}
\end{figure}

Simulations of these three realizations revealed that the models
scarcely evolved when $\sigma_{\rm gas} = 8$ or 11~km~s$^{-1}$, but
that with $\sigma_{\rm gas} = 5\;$km~s$^{-1}$ was unstable and evolved
similarly to the ``Case 2'' model shown in Figure \ref{fig.proj2}.
The changes to the properties of the coolest of our Case 3 models are
reported in Figure~\ref{fig.m3-5final}; the other two models scarcely
evolved.

It is interesting to note that the minimum value of $Q_{\rm gas} \sim
0.15\sigma_{\rm gas}$ in these models, with $\sigma_{\rm gas}$ in
km~s$^{-1}$.  Thus, the minimum $Q_{\rm gas} > 0.75$, 1.2 and 1.7 for
the three values $\sigma_{\rm gas}$ we have tried, and that only the
model with $Q_{\rm gas}<1$ at some radii was unstable.  Furthermore,
the models of \S\ref{sec.incl} for which we reduced the inclination
had $Q_{\rm gas}>1$ everywhere only when $i \sim 15^\circ$, which also
happened to be the only stable model.  A possible inference from these
results is that global instability is triggered when $Q_{\rm gas}<1$
somewhere in the disc.

Of course, $Q \ga 1$ everywhere is not a sufficient condition for
stability -- a significant halo mass is also required.  In fact,
increasing the velocity dispersion in the gas component of the no-halo
model in \S3 did not stabilize the disc; it formed a strong bar
instead as expected from the earlier simulations by \citet{Hohl71},
\citet{OP73} and analysis by \citet{Kaln78}.

\section{Conclusions}
Rotationally-supported galaxy discs have long been known to be
susceptible to various instabilities.  The classic study by
\citet{To64} determined the $Q\geq1$ criterion for axisymmetric
stability, which has not needed significant subsequent revision.
Although the mechanism for bar-forming instabilities \citep{Hohl71,
  Kaln78, sand} is now understood \citep{To81, BT08}, no general
stability criterion is known and there are galaxies, such as M33,
whose global stability remains a puzzle \citep{SSL}.

At face value, the properties of AGC~114905 measured by MP22 are
extraordinary in {\em two} respects.  Not only did they find that the
attraction from the baryonic matter, stars and gas, could alone
account for the centrifugal balance of the disc, but their
measurements also seemed to require that $Q\ll1$ over a significant
radial range.  These properties suggest that both axisymmetric and
bar-forming instabilities would be unavoidable, and our simulations
have confirmed that a disc model having these properties disrupts
violently on a dynamical time scale.

At the distance and inclination of the disc favoured by MP22, the
orbit period at their outermost measured point is $\sim 2.6\;$Gyr, or
$\sim 20\%$ of the Hubble time, although it is shorter if the galaxy
is closer and/or the inclination has been overestimated.  Such a long
dynamical time raises the possibility that the gas disc may not be
relaxed or in centrifugal balance, and calls into question the
validity of any mass modelling and conclusions from stability
properties.  However, MP22 observed an ordered flow pattern in the
gas, which does suggest that the galaxy is both in a settled dynamical
state and not subject to violently disruptive instabilities.

The standard method to stabilize a disc \citep{OP73} is to embed it in
a DM halo, and we have conducted simulations to confirm that this
galaxy is stabilized by a large fraction of dark matter.  The
mechanism by which this works for both axisymmetric and bar-forming
instabilities is to raise the epicyclic frequency, causing material in
the disc to be more strongly tied to its home radius so that
large-scale instabilities are thereby inhibited.  Our simulations
reported in \S\ref{sec.stretch} found that not only did the DM mass
fraction have to exceed the baryonic mass, but that stability was not
achieved unless $Q>1$ everywhere also, which required the velocity
dispersion of the gas to be significantly higher than that estimated
by MP22.  If the gas velocity dispersion were held at the observed
value, we found in \S\ref{sec.incl} an even larger DM mass fraction,
$\sim 3$ times the baryon mass, was needed for stability.

However, these solutions do not fit comfortably with the findings of
MP22.  They converted their observed orbital speeds in projection,
$V\sin i$, where $V$ is the speed in the plane, by adopting an
inclination $i \sim 32^\circ$ of the disc plane to the line of sight,
which they estimated from the projected shape of the gas disc.  As
reviewed in the introduction, it is difficult to estimate the true
value of $i$ when the disc is this close to face on.  Thus we
considered models having a significantly decreased value for $i$,
which increased the deprojected circular speed and allowed sufficient
dark matter to stabilize the disc.

One can also imagine that the gas velocity dispersion, $\sigma_{\rm
  gas}$, has been underestimated.  The authors of MP22 concede
(private communication) that it is hard to measure the gas velocity
dispersion in their low S/N data and the value could exceed the
$\sigma_{\rm gas} \sim 5\;$km~s$^{-1}$ they gave in their paper.
However, the outer discs of large galaxies, beyond $R_{25}$, have
velocity dispersions in the gas of 5-7~km~s$^{-1}$ \citep{Kamp93,
  Tamb09}.  This consideration leads us to prefer $\sigma_{\rm gas}
\simeq 5\;$km~s$^{-1}$ in AGC~114905, and therefore a lower value for
$i\; (\sim 15^\circ)$ is needed for this galaxy to be stable.

Our findings are consistent with almost all previous work on disc
stability, both analytic and simulations.  But the apparently
anomalous bar stability of M33, alluded to above, suggests that our
understanding of disc global stability may be incomplete.  However,
M33 is a member of the Local Group and has a substantially warped
outer gas layer.  While there is no obvious connection between these
facts and its anomalous stability, we note that neither of them
pertain to AGC~114905 and therefore our conclusions about the bar
stability of AGC~114905 may be reasonable.  Furthermore, there is
no reason to question that axisymmetric Jeans stability requires
$Q\geq1$, which alone demands an inclination of AGC~114905 much lower
than $i\sim32^\circ$.

One possible cause of an overestimated inclination could be that the
disc is, in fact, non-axisymmetric.  The sample of UDGs from which
MP22 chose AGC~114905 have low inclinations in general, and the
estimated inclination is very sensitive to mild oval distortions of
the disc.  For low inclination systems an intrinsic non-axisymmetric
shape generally biases the estimated inclination to higher values
\citep[\eg][last panel of their Fig.~3]{Li18}.

All these objects are known to have low surface density and a gas disc
that is more massive and extensive than the stellar disc, although
MP22 currently have sufficient information to determine the rotation
curve and the value of $Q$ in AGC~114905 only.  The stability of this
class of galaxies would be more difficult to understand if other UDGs
in their sample also turned out to be apparently unstable.  It is
possible that such objects are intrinsically non-axisymmetric because
of the sort of instabilities considered here.  On the other hand it
may be best not to attach much significance to models of discs having
inclinations less than 40$^\circ$, as has been traditional in rotation
curve studies.

\section*{Acknowledgements}
We thank both Filippo Fraternali and Pavel Mancera Pi\~na for valuable
comments on our results, suggestions for new tests, and for being
helpfully engaged with our project throughout.  We are also grateful
to Moti Milgrom for very useful comments on the role of inclination in
biasing the Tully-Fisher relation, and the referee for a helpful,
though tardy, report.  JAS acknowledges the continuing hospitality of
Steward Observatory.

\section*{Data availability}
The data from the simulations reported here can be made available
on request.  The simulation code can be downloaded from
{\tt http://www.physics.rutgers.edu/galaxy}


\bsp	
\label{lastpage}
\end{document}